# MHD BLOOD FLOW AND HEAT TRANSFER THROUGH AN INCLINED POROUS STENOSED ARTERY WITH VARIABLE VISCOSITY

BHAVYA TRIPATHI*[#], B. K. SHARMA*

*Department of Mathematics, BITS Pilani, Pilani Campus, Rajasthan – 333031, India,
[#]e-mail: bhaviiitr2013@gmail.com

*Abstract.* In this article, effects of heat transfer on MHD blood flow through a stenosed inclined porous artery with heat source have been investigated. The viscosity of the blood is assumed to be varying radially with hematocrit throughout the region of the artery. Governing equations have been derived by treating blood as incompressible magnetohydrodynamic (MHD) Newtonian fluid. Momentum and energy equations of the fluid flow are simplified under the assumption of mild stenosis. Homotopy perturbation method (HPM) is used to solve nonlinear differential equations for velocity and temperature profiles of the blood flow. Variation of flow rate and shear stress for different values of inclination angle and hematocrit parameter along the diseased part of artery have been plotted graphically. For having the adequate insight of the flow pattern in the diseased artery, velocity contours have been plotted for different values of the height of the stenosis and for different inclination angles of the artery.

## INTRODUCTION

In the circulatory system of our human body, artery delivers oxygenated rich blood with nutrients from the heart to each cell of the body. The abnormal elongation of arterial thickness is the first step in the formation of atherosclerosis disease. The accumulation of substances in the artery along the wall is known as stenosis, the presence of which changes the flow behavior and hemodynamic conditions of the artery [2]. As hemodynamics is directly related to overall human health, recently it has gained a serious attention of researchers, physiologists and clinical persons to study the blood flow through arteries. A body of work, in this context of arterial blood flow in the presence of stenosis, has been reported by Mekheimer and Kot [6] and Nadeem *et al.* [7] in their papers.

Magnetohydrodynamics (MHD) is the science in which we study the dynamics of electrically conducting fluids in the presence of magnetic field without considering the magnetization or polarization effects of the fluid. MHD has numerous proposed applications in bioengineering and medical sciences [9]. To analyze the flow of a biomagnetic fluid Sharma *et al.* [10] studied the effects of

___________

magnetic field on blood flow along with magnetic particles and reported that velocity of the blood flow and magnetic particles decrease as the influence of the applied magnetic field increases. Considering the non-linear model of micropolar fluid Shit and Roy [13] studied the effects of the induced magnetic field on blood flow through a constricted channel and demonstrated that increasing effects of magnetic field reduces the velocity of the blood flow at the centerline of the channel and increases the value of pressure gradient. All of these studies address the effects of magnetic field on blood flow assuming constant viscosity of the fluid flow.

However, in a real physiological system, the blood viscosity is not constant; it may vary either in hematocrit ratio or depends on temperature and pressure of the blood flow [14]. Layek *et al.* [5] analyzed the functional dependence of blood viscosity on hematocrit and suggested that viscosity increases with increase in the value of the hematocrit parameter. In the same direction, Sinha and Misra [15] presented a model of a dually stenosed artery with hematocrit-dependent viscosity.

Hyperthermia therapy is the medical treatment to treat cancer, in which body tissue is exposed to high temperatures (41–45 °C) (Habash *et al.* [4]), using external and internal heating devices. In this thermal treatment, sublethal doses of heat sensitize the cancer cells to radiation [1]. Sharma *et al.* [11–12] developed the mathematical models of magneto pulsatile blood flow through a porous medium with a heat source and radiation. They documented those more negative values of heat source parameter result in a higher temperature of the blood flow. Petrofsky *et al.* [8] examined the moisture content of the heat source on the skin blood flow response. They found in their analysis that with a set increase in skin temperature, moist heat is a better heating modality than the dry heat.

Motivated by the above analysis, this paper presents an analytical study of the effect of heat transfer on MHD blood flow through a stenosed inclined porous artery having variable viscosity and heat source. In the mathematical model of the given problem, momentum and energy equations of the blood flow are solved under the well-defined boundary conditions using homotopy perturbation method [3]. Effects of different physical parameters such as the inclination angle of the artery, porosity parameter, heat source parameter and magnetic field parameter on velocity and temperature profile of the blood flow have been plotted through graphs.

## MATHEMATICAL MODEL

Let us consider an incompressible MHD Newtonian fluid of density $\rho$ and variable viscosity $\bar{\mu}$ through a porous artery of finite length $L$. Artery having stenosis of height $\delta$ is inclined at an angle $\gamma$ from the vertical axis. The artery is assumed cylindrical in which $(\bar{r}, \bar{\theta}, \bar{z})$ represents the coordinate of a material point and $(\bar{u}, \bar{v}, \bar{w})$ are the velocity components at respective point. The blood flow is assumed to be flowing in the axial direction through cylindrically shaped non-tapered porous artery under the influence of applied magnetic field as shown in Figure 1. Throughout the region of the arterial tube density of the fluid is constant while the viscosity is assumed to be varying radially due to hematocrit variation. A case of symmetrically shaped mild stenosis is considered to convert the model in non-dimensional form.

Under these assumptions equations governing the flow under the consideration are as follows:

$$\frac{\partial \bar{u}}{\partial \bar{r}} + \frac{\bar{u}}{\bar{r}} + \frac{\partial \bar{w}}{\partial \bar{z}} = 0, \tag{1}$$

$$\rho \left[ \bar{u} \frac{\partial \bar{u}}{\partial \bar{r}} + \bar{w} \frac{\partial \bar{u}}{\partial \bar{z}} \right] = -\frac{\partial \bar{P}}{\partial \bar{r}} + \frac{\partial}{\partial \bar{r}} \left[ 2\mu(\bar{r}) \frac{\partial \bar{u}}{\partial \bar{r}} \right] + 2 \frac{\mu(\bar{r})}{\bar{r}} \left[ \frac{\partial \bar{u}}{\partial \bar{r}} - \frac{\bar{u}}{\bar{r}} \right] + \frac{\partial}{\partial \bar{z}} \left[ \mu(\bar{r}) \left( \frac{\partial \bar{u}}{\partial \bar{z}} + \frac{\partial \bar{w}}{\partial \bar{r}} \right) \right], \tag{2}$$

$$\rho \left[ \bar{u} \frac{\partial \bar{w}}{\partial \bar{r}} + \bar{w} \frac{\partial \bar{w}}{\partial \bar{z}} \right] = -\frac{\partial \bar{P}}{\partial \bar{z}} + \frac{\partial}{\partial \bar{z}} \left[ 2\mu(\bar{r}) \frac{\partial \bar{w}}{\partial \bar{z}} \right] + \frac{1}{\bar{r}} \frac{\partial}{\partial \bar{r}} \left[ \mu(\bar{r}) \bar{r} \left( \frac{\partial \bar{u}}{\partial \bar{z}} + \frac{\partial \bar{w}}{\partial \bar{r}} \right) \right] - \sigma_1 \overline{\mu_m}^2 H_0^2 \bar{w} + \rho g \alpha \left( \bar{T} - \bar{T}_0 \right) \cos \gamma - \frac{\mu(\bar{r}) \bar{w}}{k_1}, \tag{3}$$

$$\rho c_p \left[ \bar{u} \frac{\partial \bar{T}}{\partial \bar{r}} + \bar{w} \frac{\partial \bar{T}}{\partial \bar{z}} \right] = \frac{k}{\bar{r}} \frac{\partial}{\partial \bar{r}} \left( \bar{r} \frac{\partial \bar{T}}{\partial \bar{r}} \right) + \mu(\bar{r}) \left( \frac{\partial \bar{w}}{\partial \bar{r}} \right)^2 - A(\bar{T} - \bar{T}_0), \tag{4}$$

where the term $\rho g \alpha \left( \bar{T} - \bar{T}_0 \right)$ is the body force in terms of temperature, $A(\bar{T} - \bar{T}_0)$ is the expression for the external heat source, $c_p$ is the specific heat at constant pressure, $k$ is the thermal conductivity, $k_1$ is the permeability of porous medium and $\sigma_1$ is the electrical conductivity and $\mu_m$ is the magnetic permeability. $\mu(\bar{r})$ is the variable viscosity of the blood flow which is defined as:

$$\mu(\bar{r}) = \mu_0 (1 + \lambda h(\bar{r})), \tag{5}$$

where $h(\bar{r}) = H \left[ 1 - \left( \frac{\bar{r}}{d_0} \right)^m \right]$, and $H_r = \lambda H$, in which $\lambda$ is a constant having the value 2.5 and $H$ is the maximum hematocrit at the center of an artery, $m$ is the parameter that determines the exact shape of the velocity profile of blood and $H_r$ is the hematocrit parameter [15].

The geometry of the stenosis located at point $z$ with its maximum height of $\delta$ is defined as [6].

$$h(\bar{z}) = d(\bar{z})[1 - \eta(b^{n-1}(\bar{z} - a) - (\bar{z} - a)^n)] \text{ when } a \leq \bar{z} \leq a + b \quad (6)$$
$$= d(\bar{z}), \quad \text{otherwise}$$

where $d(\bar{z})$, is the radius of the tapered artery in the stenotic region with $d(\bar{z}) = d_0 + \xi\bar{z}$.

In Eq. (6), we consider $n = 2$ for symmetric stenosis case and value of $\eta$ as:

$$\eta = \frac{\delta^* n^{\frac{n}{n-1}}}{d_0 b^n (n-1)}. \quad (7)$$

We introduce the following dimensionless parameters:

$$\bar{u} = \frac{u u_0 \delta}{b}, \bar{r} = r d_0, \bar{z} = zb, \bar{w} = w u_0, \bar{h} = h d_0, \bar{P} = \frac{u_0 b \mu_0 P}{d_0^2}, \text{Re} = \frac{\rho b u_0}{\mu_0},$$

$$\Theta = \frac{(\bar{T} - \bar{T_0})}{\bar{T_0}}, \Pr = \frac{\mu c_p}{k}, Ec = \frac{u_0^2}{c_p T_0}, Z = \frac{k_1}{d_0^2}, M^2 = \frac{\sigma_1 H_0^2 d_0^2}{\mu_0}, \quad (8)$$

$$Gr = \frac{g \alpha d_0^3 \bar{T_0}}{v^2}, Q = A \frac{d_0^2}{k},$$

where Re is the Reynolds number, $Ec$ is the Eckert number, Pr is the Prandtl number and $Gr$ is the Grashof number and Brinkman number $Br = Ec \Pr$.

In the case of mild stenosis $\frac{\delta^*}{d_0} \ll 1$ and using the other two additional conditions [9] which are given as:

$$\frac{\text{Re}\delta^* n^{\frac{1}{n-1}}}{b} \ll 1, \frac{d_0 n^{\frac{1}{n-1}}}{b} \sim O(1). \quad (9)$$

Eqs. (1–4), leads the following model in non-dimensional form as:

$$\frac{\partial w}{\partial z} = 0, \quad (10)$$

$$\frac{\partial P}{\partial r} = 0, \quad (11)$$

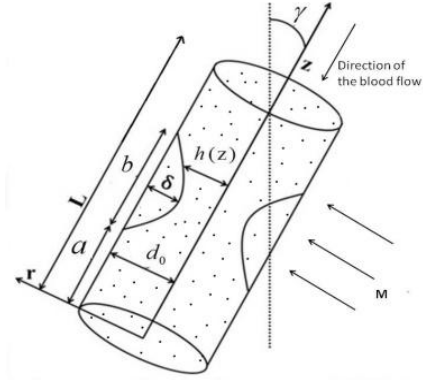

Fig. 1. The geometry of the inclined non-tapered artery.

$$\frac{\partial P}{\partial z} = \left[\frac{1}{r} + H_r\left(\frac{1}{r} - (m+1)r^{m-1}\right)\right]\frac{\partial w}{\partial r} + \left[1 + H_r(1-r^m)\frac{\partial}{\partial r}\left(\frac{\partial w}{\partial r}\right)\right] + \\ G_r\Theta\cos\gamma - w\left(M^2 + \frac{1}{Z} + \frac{H_r}{Z}(1-r^m)\right), \quad (12)$$

$$\frac{1}{r}\frac{\partial}{\partial r}\left[r\frac{\partial\Theta}{\partial r}\right] + E_c P_r\left[\frac{\partial w}{\partial r}\right]^2 - Q\Theta = 0. \quad (13)$$

The boundary conditions are as follows:

$$\frac{\partial w}{\partial r} = 0, \frac{\partial\Theta}{\partial r} = 0 \text{ at } r = 0, \quad (14)$$

$$w = 0, \Theta = 0 \quad \text{at } r = h(z), \quad (15)$$

where $h(z)$ is the geometry of the stenosis in non-dimensional form. In case of unit length artery, i.e. $d_0 = 1$, the geometry of the stenosis is given by:

$$h(z) = (1 + \xi'z)[1 - \eta_1((z - l_1) - (z - l_1)^n)] \quad \text{when } l_1 \leq z \leq l_1 + 1, \\ = 1, \quad \text{otherwise} \quad (16)$$

where $\eta_1 = \frac{\delta n^{\frac{n}{n-1}}}{(n-1)}, \delta = \frac{\delta^*}{d_0}, l_1 = \frac{a}{b}, \xi' = \frac{\xi b}{d_0}, \xi = \tan(\phi)$.

## SOLUTION OF THE PROBLEM

Now we use semi-analytical homotopy perturbation method to solve the nonlinear Eqs. (10–13) under the given boundary conditions by Eqs. (14–15). Therefore, we first formulate the following homotopy for momentum Eq. (12) of the blood flow:

$$H(q,w) = q\left[L(w) + H_r(\frac{1}{r} - (m+1)r^{m-1})\frac{\partial w}{\partial r} + H_r(1-r^m)\frac{\partial}{\partial r}(\frac{\partial w}{\partial r}) - \frac{\partial p}{\partial z}\right]$$
$$+(1-q)[L(w) - L(w_0)] - q\left[w(M^2 + \frac{1}{Z} + H_r\frac{(1-r^m)}{Z})\right] + \quad (17)$$
$$q\cos\gamma(G_r\Theta),$$

where $q \in [0,1]$ is the embedding parameter and $L(w)$ is linear operator defined as:

$$L(w) = \frac{1}{r}\left(\frac{\partial}{\partial r}\left(r\frac{\partial w}{\partial r}\right)\right).$$

Similarly, homotopy for the energy Eq. (13) is formulated as:

$$H(q,\Theta) = (1-q)\left[L(\Theta) - L(\Theta_0)\right] + q\left[L(\Theta) + E_c P_r\left(\frac{\partial w}{\partial r}\right)^2 - Q\Theta\right]. \quad (18)$$

The initial guesses used to solve these homotopies Eqs. (17–18) are given by:

$$w_{10} = \frac{(r^2 - h^2)}{4}\left(M^2 + \frac{1}{Z}\right)\left(\frac{\partial p_0}{\partial z}\right), \quad (19)$$

$$\Theta_{10} = \frac{(r^2 - h^2)}{4}. \quad (20)$$

Now, the dependent variables $w(r,q)$ and $\Theta(r,q)$ can be decomposed as:

$$w(r,q) = w_0 + qw_1 + q^2 w_2 + O(q^3), \quad (21)$$
$$\Theta(r,q) = \Theta_0 + q\Theta_1 + q^2\Theta_2 + O(q^3). \quad (22)$$

Now, substitute the series expansion of the above variables from Eqs. (21–22) into Eq. (17) and Eq. (18) compare the coefficients of $q^0$, $q^1$ and $q^2$ we get:

$$q^0 : L(w_0) - L(w_{10}) = 0 \Rightarrow w_0 = w_{10} = \frac{(r^2 - h^2)}{4}\left(M^2 + \frac{1}{Z}\right)\left(\frac{\partial p_0}{\partial z}\right), \quad (23)$$

$$q^1 : L(w_1) = -L(w_0) - H_r\left(\frac{1}{r} - (m+1)r^{(m-1)}\right)\frac{\partial w_0}{\partial r} - \cos\gamma\left(G_r\Theta_0\right) + \frac{\partial p_0}{\partial z} -$$
$$H_r\left(1 - r^m\right)\frac{\partial}{\partial r}\left(\frac{\partial w_0}{\partial r}\right) + w_0\left(M^2 + \frac{1}{Z} + H_r\frac{(1-r^m)}{Z}\right), \quad (24)$$

$$L(w_2) = -H_r\left(\frac{1}{r} - (m+1)r^{(m-1)}\right)\frac{\partial w_1}{\partial r} + w_1\left(M^2 + \frac{1}{Z} + H_r\frac{(1-r^m)}{Z}\right) +$$
$$\frac{\partial p_1}{\partial z} - H_r(1-r^m)\frac{\partial}{\partial r}\left(\frac{\partial w_1}{\partial r}\right) - \cos\gamma\left(G_r\Theta_1\right). \quad (25)$$

Now in Eq. (18), compare the coefficients of $q^0$, $q^1$ and $q^2$ respectively:

$$q^0 : L(\Theta_0) - L(\Theta_{10}) = 0 \Rightarrow \Theta_0 = \Theta_{10} = \frac{(r^2 - h^2)}{4}, \quad (26)$$

$$q^1 : L(\Theta_1) = -L(\Theta_0) - E_c P_r \left(\frac{\partial w_0}{\partial r}\right)^2 + Q\Theta_0, \quad (27)$$

$$q^2 : L(\Theta_2) = -2E_c P_r \left(\frac{\partial w_0}{\partial r}\right)\left(\frac{\partial w_1}{\partial r}\right) + Q\Theta_1. \quad (28)$$

So with the help of initial guesses $w_{10}$, $\Theta_{10}$ from Eq. (23) and Eq. (26) and using the definition of linear operators $L(w_0)$ and $L(\theta_0)$, in Eq. (24) and Eq. (27), we get the expressions for $w_1$ and $\theta_1$ as:

$$w_1 = \frac{(r^2 - h^2)}{4}\left(\frac{\partial p_0}{\partial z}\right)\left(1 - (M^2 + \frac{1}{Z})\right) - \frac{\cos\gamma}{64}\left(r^4 + 3h^4 - 4r^2h^2\right)G_r +$$

$$\frac{H_r}{4Z}\left(\frac{\partial p_0}{\partial z}\right)\left(M^2 + \frac{1}{Z}\right)\left(\frac{r^4 + 3h^4}{16} + \frac{(h^{m+4} - r^{m+4})}{(m+4)^2} + \frac{h^2 r^{m+2} - h^{m+4}}{(m+2)^2}\right) -$$

$$\frac{H_r}{4Z}\left(\frac{\partial p_0}{\partial z}\right)\left(M^2 + \frac{1}{Z}\right)\left(\frac{r^2 h^2}{4}\right) + \left(\frac{\partial p_0}{\partial z}\right)\left(M^2 + \frac{1}{Z}\right)^2\left(\frac{r^4}{16} + \frac{3h^4}{16} - \frac{h^2 r^2}{4}\right) -$$

$$\frac{H_r}{2}\left(\frac{r^2 - h^2}{2} + \frac{h^{m+2} - r^{m+2}}{m+2}\right)\left(\frac{\partial p_0}{\partial z}\right)\left(M^2 + \frac{1}{Z}\right), \quad (29)$$

$$\Theta_1 = -E_c P_r \frac{(r^4 - h^4)}{64}\left(\frac{\partial p_0}{\partial z}\right)^2\left(M^2 + \frac{1}{Z}\right)^2 + Q\left(\frac{r^4}{16} + \frac{3h^4}{16} - \frac{r^2 h^2}{4}\right) -$$

$$\left(\frac{r^2 - h^2}{4}\right). \quad (30)$$

Expressions for $w_2$ and $\theta_2$ have been calculated using MATLAB 2015a by substituting the values of $w_1$, $\theta_1$ in Eq. (25) and Eq. (28). Therefore, the final expressions for velocity $w(r, z)$ and temperature $\theta(r, z)$ are obtained by substituting the values of $w_1$, $w_2$, $\theta_1$ and $\theta_2$ in Eq. (21) and Eq. (22) respectively.

## RESULTS AND DISCUSSIONS

In this section graphical results have been displayed for velocity, temperature and wall shear stress profiles of the blood flow through the stenosed artery for different quantities of interest. Default values which have been used to graphically analyze the effectiveness of the model are as follows:

$d_0 = 1, M = 1.5, h(z) = 0.9, a = 0.25, b = 1, \delta = 0.1, Z = 0.3, n = 2, G_r = 2, z = 0.5,$
$\gamma = \frac{\pi}{6}, H_r = 1, t = 0.$

Figure 2, displays the effect of magnetic field parameter ($M$) on velocity profile of the blood flow. It can be easily seen from the figure 2 that the velocity profile of the blood flow decreases as values of the magnetic field parameter increase. It happens because of the fact that when blood flows under the influence of magnetic field, the action of magnetization applies a rotational motion on the charged particles of the blood flow. The continuous rotational motion of the charged particles causes red blood cells to be more suspended in the blood plasma and this effect increases the internal viscosity of the blood flow. Increased viscosity causes rise to a resistive type force called the Lorentz force. This force

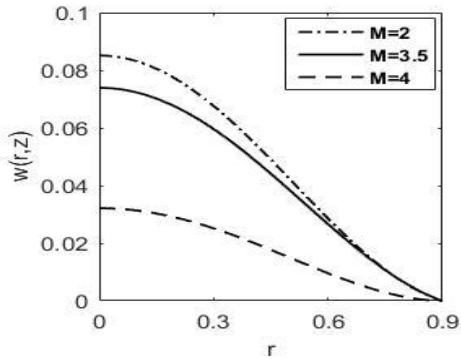
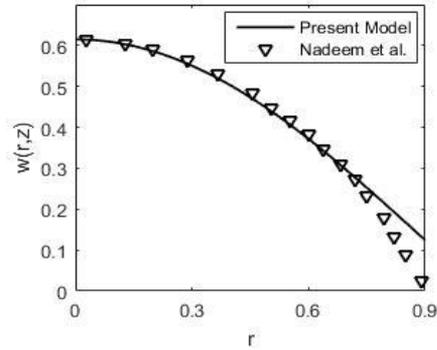

Fig. 2. Velocity variation with *M*.  Fig. 3. Comparison between present study and the result presented by Nadeem *et al*. [7].

opposes the motion of blood particles, which reduces the velocity of the blood flow. This effect of magnetic field on velocity profile of the blood flow follows in the same manner as proposed by the authors Sharma *et al*. [10].

Figure 3 shows the comparison between the present result and the result presented by Nadeem *et al.* [7] Comparison result shows a good agreement between the present result and the result presented by Nadeem *et al.* [7] for the given values $M = 0.2, Z = 0.7, Q = 0, \delta = 0.1, \gamma = \pi, G_r = 0.$

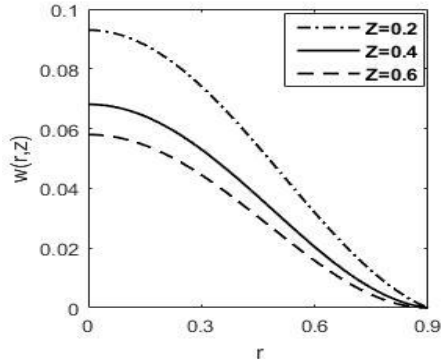 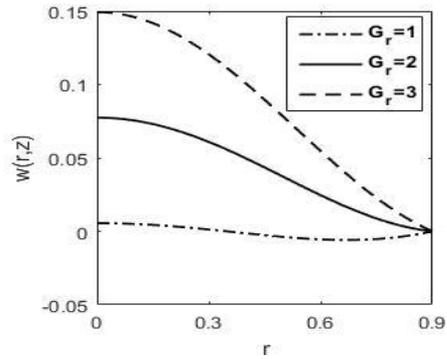

Fig. 4. Velocity variation with *Z*.   Fig. 5. Velocity variation with $G_r$.

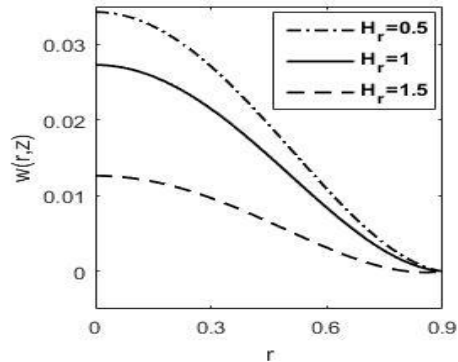 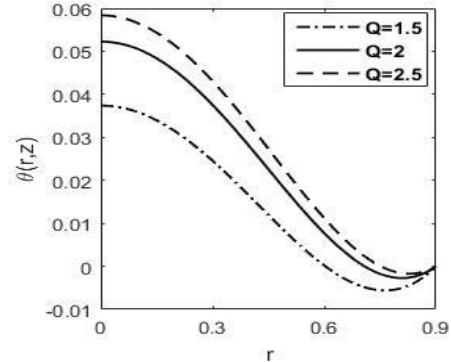

Fig. 6. Velocity variation with $H_r$.   Fig. 7. Temperature variation with *Q*.

Figure 4 shows the effects of porosity parameter $(Z)$ on the velocity profile of the blood flow. It is clear from the figure that velocity profile decreases with increasing values of the porosity parameter. Therefore, for a particular value of porosity parameter, velocity attains its maximum value at the middle of the artery and it starts decreasing towards the wall of the artery. Velocity profile with porosity parameter shows this behavior may be because, when a fraction of the voids volume over the total volume increases, it can become increasingly difficult for fluid to move from one place to another place in the artery. Hence the velocity of the blood flow through an inclined porous artery decreases with increasing values of the porosity parameter.

In figure 5, the velocity profile of the blood flow is plotted for different values of Grashof number $(G_r)$. It is clear from the figure 5 that as values of the Grashof number changes from 1 to 3, the velocity profile of the blood flow increases and this happens due to increased Boussinesq source terms. Grashof number indicates the relative effect of the thermal buoyancy force to the viscous hydrodynamic force in the boundary layer. Therefore, a rise in the velocity of the blood flow is observed due to the enhancement of thermal buoyancy force as values of the Grashof number increase.

Figure 6 illustrates the distribution of velocity profile of the blood flow for different values of the hematocrit parameter. The figure shows that the velocity profile of the blood flow decreases with increasing values of hematocrit parameter. In blood, hematocrit is the volume percentage of red blood cells and in the artery as the number of red blood cells increases in the volume, the density of blood flow increases relatively. Increased density of blood slows down the flow of blood and this causes the decreased velocity of the blood flow. The figure further reveals that velocity of the blood flow follows the parabolic profile such that for any particular value of the hematocrit parameter velocity attains maximum value at the center and it starts decreasing as we move towards the stenotic wall of the artery.

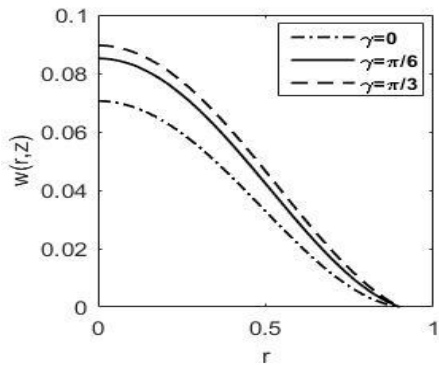 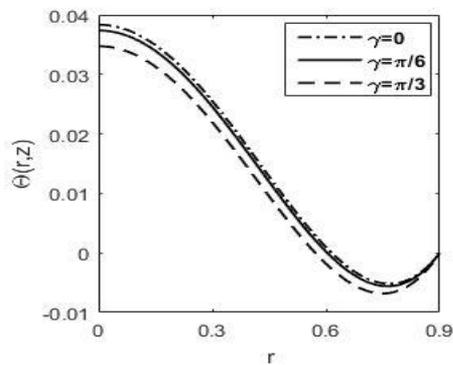

Fig. 8. Velocity variation with γ.                Fig. 9. Temperature variation with γ.

Figure 7 marks the variation of the temperature profile of the blood flow for different values of heat source parameter $(Q)$. It can be marked that within the blood, radiation acts as a heat source. Hence, increasing values of radiation dosage would directly increase the temperature profile of the blood flow. In the pathological state, this type of thermal therapy is very much used to expose body tissues and cancerous tumor to high temperature. The result of which kills cancer cells associated with tumors with minimal injury to normal tissues. It is observed through the figure that for a particular value of the heat source parameter, temperature profile of the blood flow achieves its maximum value at the center of the artery and it starts decreasing as we move towards the stenotic wall.

Figures 8 and 9 illustrate the distribution of velocity and temperature profiles of the blood flow with different inclination angles of the artery, respectively. Figure 8 displays that as values of the inclination angle (made by the artery from the vertical axes) increase velocity profile of the blood flow also increases.

Increased values of inclination angle have reverse effects on the temperature profile of the blood flow as figure 9 displays that as the inclination angle of the artery changes from $0$ to $\frac{\pi}{3}$, the temperature profile of the blood flow decreases.

Expression for the wall shear at the maximum height of the stenosis is given by:

$$\bar{\tau}_s = \left[\frac{\partial w}{\partial r}\right]_{h=1-\delta}. \qquad (31)$$

Flow rate of the blood flow is calculated as:

$$Q_w = 2\pi R^2 \int_0^R w(r)\,dr. \qquad (32)$$

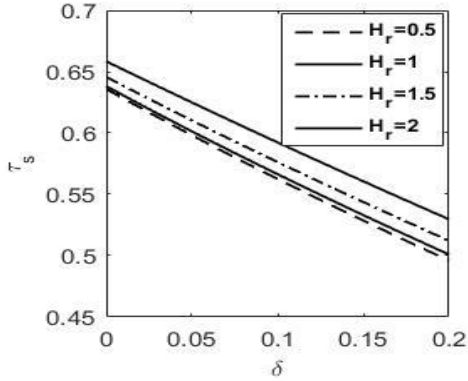 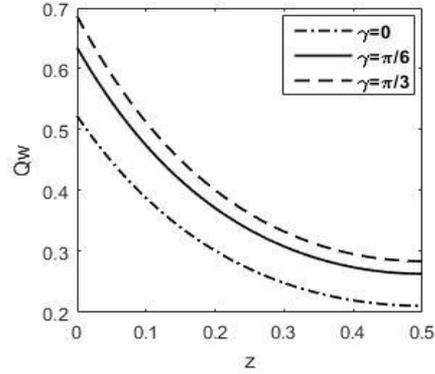

Fig. 10. Shear Stress variation with $H_r$.   Fig. 11. Flow rate variation with $\gamma$.

Figure 10 displays the variation of wall shear stress at stenosis throat with $H_r$ for different values of the hematocrit parameter. The figure illustrates that as values of the hematocrit parameter increase from $0.5$ to $2$ wall shear stress at stenosis throat also increases. Figure 11 shows that flow rate of the blood flow along with the axial distance for different inclination angles of the artery. The figure reveals that as values of the inclination angle increasingly change from $0$ to $\frac{\pi}{3}$, the flow rate of the blood flow near stenosis also increases.

## CONCLUSION

In the article the hemodynamics of MHD blood flow having variable viscosity through a stenosed, inclined arterial segment with heat source has been investigated. The study discusses the situation when the lumen of an arterial segment converts into a porous medium due to deposition of fatty substances. Flow dynamics has been analysed by the momentum and enery equation of the flow. Homotopy perturbation technique is used to get the analytical solutions for velocity and temperature equations of the blood flow. The significant findings of this paper are summarized as follows:

- The velocity of the blood flow increases as values of the magnetic field parameter increase and this happens due to the Lorentz force which opposes the motion of the blood flow in the artery. The given result can be very much useful to control blood flow during the surgical process.
- The velocity profile of the blood flow increases as values of the inclination angle made by the artery increase while it shows the reverse effects on the temperature profile of the blood flow.
- The velocity of the blood flow decreases as values of the hematocrit parameter increase. This happens due to increased density of the red blood cell inside the artery which in result slows down the flow.
- Heat source parameter brings out an effective change in the temperature profile of the blood flow. The temperature profile of the blood flow increases as values of the heat source parameter increase. This result is very much useful bearing on the therapeutic procedure of hyperthermia, particularly in understanding/regulating the blood flow.
- The shear stress at the wall of the stenosis increases as values of the hematocrit parameter increase. The flow rate of the blood flow increases as values of the inclination angle made by the artery increase.

*Acknowledge*: Authors are sincerely thankful to the Department of Science and Technology, Government of India (SR/FST/MSI-090/2013(C)) for their financial support.